\documentclass[preprint,12pt,number]{elsarticle}
\usepackage{amssymb}
\usepackage{xurl}
\usepackage{comment}
\usepackage{multirow}
\usepackage{tabularx}
\usepackage{amsmath}
\usepackage{enumitem}
\usepackage{xcolor}
\usepackage{wasysym}
\usepackage{listings}
\usepackage{courier}
\usepackage{caption}
\usepackage{float}
\usepackage[caption = false]{subfig}
\usepackage{graphicx}
\usepackage{tikz}
\usepackage{array}
\usetikzlibrary{positioning, shapes, arrows}
\usepackage{hyperref}
\hypersetup{
    colorlinks=true,
    anchorcolor=black,
    citecolor=blue,
    linkcolor=blue,
    filecolor=blue,      
    urlcolor=blue,
}
\usepackage{xurl}
\usepackage{longtable}

\journal{}
\begin{document}
\begin{frontmatter}

\title{Classification and taxonomy of mobile application usability issues}

\author[inst1]{Paweł Weichbroth*}
\affiliation[inst1]{organization={Gdansk University of Technology, Faculty of Electronics, Telecommunications and Informatics, Department of Software Engineering},
            city={Gdansk},
            country={Poland}\\ * Corresponding Author: pawel.weichbroth@pg.edu.pl}
  
\begin{abstract}
Despite years of research on testing the usability of mobile applications, our understanding of the issues their users experience still remains fragmented and underexplored. 
While most earlier studies has provided interesting insights, they have varying limitations in methodology, input diversity, and depth of analysis. 
On the contrary, this study employs a triangulation strategy, using two research methods (systematic literature review and interview) and two data sources (scholarly literature and expert knowledge) to explore the traits underlying usability issues.
Our study contributes to the field of human–computer interaction (HCI) by presenting a catalog of 16 usability issue categories, enriched with corresponding keywords and extended into a taxonomy, as well as a novel three-tier app–user–resource (AUR) classification system.
At the first app level, usability issues arise from user interface design, as well as from efficiency, errors, and operability. 
At the second user level, they influence cognitive load, effectiveness, ease of use, learnability, memorability, and understandability. 
At the third resource level, usability issues stem from network quality and hardware, such as battery life, CPU speed, physical device button size and availability, RAM capacity, and screen size. The root cause of the usability issues is the user interface design.
Detailed findings and takeaways for both researchers and practitioners are also discussed. 
Further research could focus on developing a measurement model for the identified variables to confirm the direction and strength of their relationships with perceived usability. Software vendors can also benefit by updating existing quality assurance programs, reviews and audits tools, as well as testing checklists.
\end{abstract}


\end{frontmatter}

\section{Introduction}
\label{introduction}
Mobile applications (apps) have become an integral part of daily life by instantly connecting people and businesses around the world~\cite{rakshit2021mobile}. They have surpassed desktop applications as the primary source of information~\cite{Tiwari2023}, communication~\cite{Statista-2025a} and entertainment~\cite{Hart2025}. Currently, there are 8.93 million smartphone apps worldwide, with 3.553 million apps in the Google Play Store and 1.642 million in the Apple App Store, and the average smartphone user has 40 apps installed on their mobile device~\cite{Turner2025}. The importance of mobile applications to the global economy is clear, as this market was valued at \$333.93 billion in 2025 and is expected to reach \$745.36 billion by 2030~\cite{Morder2025}.

Although mobile applications appear less complex than their desktop counterparts at first glance~\cite{kaur2022systematic}, yet new challenges immediately emerged regarding usability~\cite{elguera2019elderly, namoun2024predicting}, among many others~\cite{lacerda2018systematic}. Obviously, mobile devices are handy and powerful, but they have several shortcomings. 
Despite the modern trend toward larger screens~\cite{MiniMicroLED2024,fbi2025}, mobile devices' true convenience and portability still come from their compact size. However, small size limits how much information can be displayed at once, especially compared to the spacious screens of laptops and desktop computers.

Another limitation is the lack of keyboard and mouse, which are replaced by a touchscreen. With the small screen becoming the default input and output, typing proficiently on a tiny virtual keyboard becomes difficult, and accidentally touching the wrong target becomes easy. Although gestures offer users a new way to interact that mirrors their real-world experience~\cite{rico2011gesture}, they also raise issues, unknown in traditional input methods.

These limitations have shaped the design of mobile applications~\cite{alshammare2025revealing}.
Bearing in mind the aforementioned obstacles, testing the design on various devices with different resolutions and operating systems becomes imperative to ensure proper functionality under all conditions~\cite{zein2016systematic}. 
Nevertheless, testing mobile applications presents its own set of challenges~\cite{kirubakaran2013mobile,arif2019mobile}. The problem is not how to perform the testing, but rather what to test.

For instance, over the past 15 years, users have added nearly 166 million online reviews for Instagram, including both the Android~\cite{Instagram-GP} and iOS versions; this means that, on average, more than 30,000 reviews have been added daily. This may seem surprising knowing that a variety of usability testing and evaluation methods and tools have been already developed in recent years~\cite{tramontana2019automated}.

Although researchers have devoted considerable attention to usability, earlier studies have provided valuable insights, they have varying limitations in methodology, input diversity, and depth of analysis. Such results can be questioned due to the narrowed credibility.
This perspective is also confirmed by Huang and Benyoucef~\cite{huang2023systematic}, who state that usability is still not well understood.
This suggest that mobile usability landscape remains an underexplored topic.

Furthermore, the reported issues are quite diverse, suggesting that they are related to the context of use or to specific application features. Such findings can be questioned due to limitations in generalizability.
This finding is also raised by Weichbroth~\cite{weichbroth2025usability}, who 
through the review of recent studies, demonstrated the diversity of the results documented over the last 20 years. This suggests that our understanding of usability traits remains fragmented. 

In our opinion, these arguments emphasize the need for comprehensive research investigating the underlying issues of mobile application usability. 
In this vein, our study relies on a triangulation approach with regard to both research methods and data sources. Our study advances the previous studies by offering three key contributions:
\begin{itemize}
    \item It presents and discusses a catalog of sixteen categories of mobile usability issues, suggesting that user interface design is the root cause of these issues and indicating that usability is associated with a variety of other qualities.
    \item It introduces a novel three-tier app-user-resource (AUR) classification system of usability issues, at the same time emphasizing an inter-related dependencies between these units.
    \item It provides taxonomy, a controlled vocabulary organized into a hierarchical structure consisting of usability issue categories and corresponding keywords. 
\end{itemize}


The remainder of the paper is structured as follows. 
Section~\ref{sec:2} discusses the theoretical background.
Section~\ref{sec:3} presents the research methodology applied. 
Section~\ref{sec:study1-slr} outlines the assumptions and results of a systematic literature review.
Section~\ref{sec:study2-interview} shows the settings and findings of an expert survey.
Section~\ref{sec:synthesis-generalization} discusses internal and external validity in terms of result synthesis.
Section~\ref{sec:final-results} introduces the catalog of mobile usability issues, and their taxonomy, including a controlled vocabulary.
Section~\ref{sec:discussion} outlines the theoretical and practical implications, study limitations along with mitigation countermeasures undertaken.
Section~\ref{sec:conclusions} concludes the paper and puts forward directions for future research.

\section{Background}
\label{sec:2}
While usability in the context of mobile applications has received several definitions over the years, the most accepted in light of recent research is the one formulated in ISO 9241-11 \cite{weichbroth2020usability}, which states that usability is the "extent to which a system, product, or service can be used by specified users to achieve specified goals with effectiveness, efficiency, and satisfaction in a specified context of use" \cite{ISO9242-11}. In this study, for the sake of methodological clarity, usability is understood in the same way.

While ISO 9241-11 is the most commonly adopted usability definition in studies of mobile applications~\cite{weichbroth2020usability}, it is a rigid view that points to three attributes: effectiveness, efficiency, and satisfaction. 
In addition, ISO 9241-11:2018 defines these notions in a straightforward manner~\cite{ISO9242-11}.
Effectiveness refers to the accuracy and completeness with which users achieve specified goals. Efficiency refers to the resources used in relation to the results achieved. Satisfaction refers to the extent to which a user's physical, cognitive, and emotional responses resulting from the use of a mobile application meet their needs and expectations.
The most recent research on usability testing that employs ISO 9241-11 also reports usability issues in terms of these three attributes~\cite{moumane2016usability,fathiyyah2022usability}.

In 2013, Harrison et al. introduced the PACMAD (People at the Center of Mobile Application Development) usability model~\cite{harrison2013usability}. It is now the most common approach in the field of mobile usability testing due to its versatility, an ability to capture the majority of user's attitudes and perceptions.
The PACMAD incorporates the ISO 9241-11 attributes and includes four additional attributes. 
First, Learnability refers to the ease with which a user can become proficient in using an application. It typically reflects the amount of time and effort required for a person to use the application effectively.
Second, Memorability refers to a user’s ability to retain how to use an application effectively after a period of not using it.
Third, Errors refer to how well a user can complete desired tasks without making mistakes.
Four, Cognitive load, representing the main contribution of the PACMAD model, refers to the amount of mental effort required from the user to operate the application. 

After the successful inception and widespread adoption of PACMAD, Ammar~\cite{ammar2019usability}, in 2019, proposed a usability model for mobile applications generated using a model-driven approach. This approach specifically emphasized the importance of Understandability and Operability. Here, the later concerns the capability of the mobile application to allow users to operate and control it. 
Four years later, in 2023, Huang and Benyoucef~\cite{huang2023systematic} pointed out that Easy of Use is also a well-recognized attribute. 
Note that Understandability and Ease of use have similar meanings. However, the latter relates to the overall logic and content presented by the user interface, while the former refers to the amount of effort a user needs to expend to interact with an application.
These studies, along with others, laid the foundation for the development of the PACMAD+3 model~\cite{weichbroth2024usability}, which incorporates all ten of the aforementioned attributes.

Typically, one way to study usability is from the perspective of the issues users encounter \cite{ali2022mobile}. Therefore, note that the evaluation of usability of mobile applications should be considered in the specific context of their use, which means that its observed attributes by the user may vary, taking into account the external (user-independent) factors such as current location, time of day, weather, quality of network connection, smartphone manufacturer, as well as internal (user-dependent) factors such as movement velocity, ongoing physical activity, or cognitive abilities. On the other hand, the basis for evaluating the methods themselves is usually the number of issues found during testing \cite{kjeldskov2004new}.

As can be seen, these models were developed incrementally and sequentially, seemingly gaining recognition in an ongoing stream of research devoted to usability testing.
However, an intriguing question emerges: Are these the state-of-the-art models sufficient for identifying and addressing usability issues? To answer this question, it is necessary to determine which usability issues have been documented thus far and what experts from the mobile application development industry have to say. Furthermore, it is also essential to consider the reviews from mobile application users themselves, as their feedback ultimately matters most. 


\section{Methodology}
\label{sec:3}
The goal of this research is to identify the usability issues of mobile applications. To achieve it, we used a triangulation approach, adopting principles presented by Noble and Heale~\cite{noble2019triangulation}.
Triangulation is applied when the field of study is difficult, demanding, or contentious~\cite{turner2009triangulation}, all of which describe presence research. In addition, triangulation enhances validity and credibility since findings from one research can be cross-confirmed by the others~\cite{ball2000putting}.

To design the triangulation-based research, the process of linking laid the methodological foundations. Turner et al. define linking as the process of combining multiple research strategies within a study to achieve its goal~\cite{turner2017research}. The linking process validates the original concepts~\cite{morse2001qualitative}, as well as enables the formulation of higher-level concepts~\cite{wafiqoh2019reflective}.

Based on the Denzin~\cite{denzin2012triangulation} classification, the employed methodology relied on methods and data triangulation. 
The former refers to the use of various methods for collecting and analyzing data, while the latter involves corroborating evidence from different sources~\cite{joo2011adoption}.

\begin{figure}[H]
    \centering
    \includegraphics[width=1\linewidth]{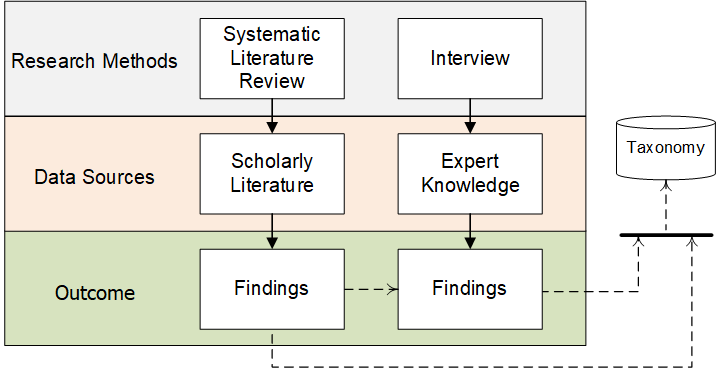}
    \caption{The triangulation framework of the current study.}
    \label{fig:methodology-framework}
\end{figure}

As can be seen in Figure~\ref{fig:methodology-framework}, our study employed two research methods and their respective data sources in the following order.

\begin{enumerate}
    \item A Systematic Literature Review (SLR) on scholarly literature was performed to extract, identify, and evaluate all relevant concepts. This is a well-recognized strategy supported by general well-established guidelines~\cite{okoli2015guide}, as well as by other studies devoted to the frontiers of information systems~\cite{carrera2022conduct}. 
    
    \item An interview was conducted to collect expert knowledge and obtain up-to-date, in-depth insights into professional experiences, as such information cannot be captured through other research methods. Numerous studies on professional users’ perceptions of human–computer interaction have frequently employed methodological approaches such as interviews~\cite{park2013developing,zimmermann2020password}.

\end{enumerate}

Note that Figure~\ref{fig:methodology-framework} also shows the data flow, indicated by a dotted line. Specifically, the SLR findings were used to organize and categorize the codes extracted from the qualitative data submitted by the experts. Next, based on the findings from these two studies, we extracted and compiled a dataset of keywords for each category to assemble a taxonomy in the form of a hierarchical controlled vocabulary, using the previously developed three-tier classification system.

\section{Study \#1. Systematic Literature Review}
\label{sec:study1-slr}
The first study was systematic literature review (SLR). 
We used the PRISMA 2020 framework in our review due to its transparency, reproducibility, and scientific rigor in research synthesis. According to Page et al.~\cite{page2021prisma}, the framework optimizes documentation at every stage, from identifying studies to including them, by providing clear guidelines that minimize bias and error.
Note that mobile application usability testing has become a popular area of research, given the dynamic development of mobile technologies and their wide range of applications. The PRISMA approach will enable us to systematically organize the processes of collecting, selecting, and evaluating information, yielding reliable results that contribute to the development of knowledge in this emerging field.

\subsection{Research questions definition}
To identify the most challenging attributes of mobile applications submitted for user testing, one of the possible approaches is to collect, analyze and synthesize the information regarding reported usability issues, so far documented in the current body of scientific literature. Having said that, in the current review, our aim is to answer the following research question (RQ):

\begin{itemize}
    \item What are the usability issues associated with mobile applications?
\end{itemize}

\subsection{Keywords extraction}
From the above research question, three keywords were extracted, namely: \textit{mobile}, \textit{usability}, and \textit{issue}. These words combined with the AND operator will serve as input to the online search engines to select the valid documents. 

\subsection{Time Frames and Data Source}
Between August 2024 and May 2025, a search was performed using the Google Scholar database, which returned about 767k results at the beginning and about 900k results at the end.

\subsection{Data Sample Development}
The process of developing data samples involved four steps:
\begin{enumerate}
    \item Collect data. The aim of this step is to gather all relevant, high-quality studies on the formulated research question using a transparent, reproducible approach.
    \item Formulate and apply inclusion criteria. These criteria must clearly define which studies are relevant and should be included in the review. Focusing the review on high-quality, applicable evidence reduces bias and improves the reliability of the findings.
    \item Formulate and apply exclusion criteria. The goal is to eliminate studies that are irrelevant, do not meet quality standards, or fall outside the scope of the research question.  Clearly defining what should be excluded contributes to the rigor, transparency, and reproducibility of the review process.
    \item Extract data. This involves the organized collection of relevant information from the input to address the research question. Extracting the documented issues from each study creates a dataset that can be analyzed and synthesized qualitatively to provide a reliable answer to the research question.
\end{enumerate}

Each step is explained in detail below, along with a brief quantitative summary.
\subsubsection{Step \#1: Data Collection}
In the first step, information including the publication title and URL was copied from the Google Scholar website into the online spreadsheet. This process was repeated until the last available page, resulting in a total of 1004 records collected. 
Note that for records marked as \textit{Citation}, only the title of the article was available. A total of five records were identified as citations and were excluded from further analysis. Therefore, the input contained 999 documents. 

\subsubsection{Step \#2: Inclusion Criteria}
In the second step, the titles of the papers were manually retrieved to verify the inclusion criteria (IC), which were defined as follows:
\begin{itemize}
    \item[IC1] The record is not a duplicate.
    \item[IC2] The title of the paper is in English.
    \item[IC3] Paper include either of the following wordings: mobile, usability.
    \item[IC4] The full text of the paper is available to the author.
    \item[IC5] Paper is a full-length article, review, or conference proceedings.
\end{itemize}

A binary scale was used to label the documents, where 1 means the criterion is met and 0 means the opposite. 
Considering the first inclusion criterion, 146 records were duplicates and were hence excluded from further analysis. Therefore, the input consisted of 853 unique papers. 
The next check of the IC2 revealed that the titles of 3 records were not in English. 
On the third pass (IC3), each title was checked for the presence of both the words \textit{mobile} and \textit{usability}. 
Note that this criterion was relaxed from the search string that indicated the presence of the third word (\textit{issue}) in order to cover a wider range of documents. Nevertheless, 264 records were excluded from the volume. 
In the fourth pass (IC4), we checked the availability of each paper using the licensed access provided by the university that is affiliated with the author. In total of 68 records were unavailable and eventually excluded. 
In the fifth pass (IC5), the source title of each paper was examined to determine its type.
Finally, after applying all inclusion criteria, the input consisted of 483 records.

\subsubsection{Step \#3. Exclusion Criteria}
The third step was to verify the exclusion criteria (EC), defined in the following way: 

\begin{itemize}
    \item[EC1] The abstract of the paper is written in a language other than English. 
    \item[EC2] Paper is written in a language other than English.
    \item[EC3] Paper is a conference keynotes, books, extended abstracts, short articles (under two pages).
    \item[EC4] Paper is an erratum, note, or editorial.
    \item[EC5] Paper has been retracted.
    \item[EC6] The research context is defined for a specific group of users or for other mobile solutions (e.g. websites).
\end{itemize}

A binary scale was used to label the documents, where 1 means the criterion is met and 0 means the opposite. 
With regard to the first and second exclusion criteria, 2 and 1 papers, respectively, were excluded from further analysis.
In the second pass, the content of each paper was examined to determine its length (EC3) and structure (EC4). After careful screening, 1 paper was excluded based on the third exclusion criterion. One paper was retracted by the publisher. 
At this point, the input included 478 records. On the third pass (EC6), the paper's abstract was subject of careful reading. Ultimately, 87 were excluded as the research targeted mobile terminals, Internet portals, mobile websites, or was devoted to unspecified mobile applications, among other reasons.
Finally, after applying all exclusion criteria, the input consisted of 391 records.

\subsubsection{Step \#4. Data Extraction}
Each paper was carefully read and analyzed to ensure that the extracted data accurately reflected its contribution to the current research topic. More specifically, we paid attention to the research methodology used and then to the reported findings, which must be directly related to the research objectives.

Beyond the aforementioned formulated exclusion criteria, we also excluded the studies involving people with disabilities, such as those presented by Leporini et al. \cite{leporini2012interacting}, as well as older adults (e.g. Wong et al. \cite{wong2018usability}), because in such cases the mobile applications are typically tailored to the needs of these, as well as other specific groups of users. In addition, usability issues must be identified through empirical testing of a particular mobile application, or through the systematic literature review, as we aim to document and synthesize only evidence-based findings.

Moreover, considering a detailed analysis of the identified usability issues reported in a particular paper, context specific issues were not taken into account. For example, while we found the study by Kjeldskov et al. \cite{kjeldskov2005evaluating} to be a valuable source of information, however there are a few context-specific issues reported such as maps ("issues related to how the user interprets and uses maps in conjunction with the textual information"), or knowledge about city ("issues related to high requirements for user’s knowledge about the city in which they are interacting with the system"), which are not generic by nature, and thus cannot be generalized. 

Eventually, based on these qualitative criteria, 61 studies were identified that discussed usability issues related to mobile applications. Interestingly, the most numerous group of studies concerned healthcare (\#26), while the remaining involved learning (\#7), general purpose (\#5), e-commerce (\#4) and tourism (\#4). In summary, the reported findings from each paper were independently copied into the external spreadsheet and served as input for the next stage of analysis, with a total amount of 2676 words.

\subsection{Data Analysis}
We used VOSviewer \cite{VOS-2025} to construct and visualize the the term network co-occurrence from the extracted dataset. This tool uses a layout algorithm to display the spatial relationships among the input items based on their similarity or the frequency of connection. By default, the outcome is an interactive network map, where items are represented as nodes and their relationships as lines, revealing clusters, patterns, and structures within the data.

\begin{figure}[!h]
    \centering
    \hspace*{-1.0in}
    \includegraphics[width=1.29\linewidth]{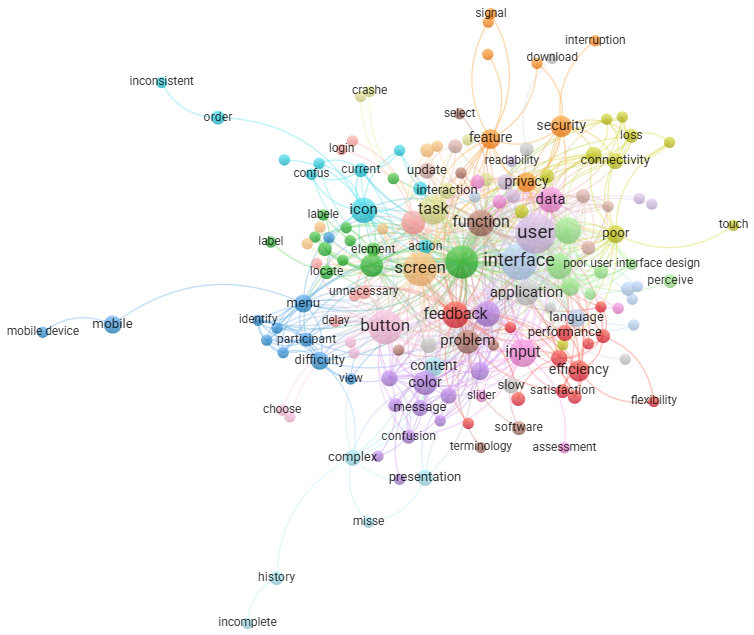}
    \caption{The network of terms used to describe the mobile usability issues.}
    \label{fig:network-usability-issues}
\end{figure}

Figure~\ref{fig:network-usability-issues} shows the network of mobile usability issues, extracted from the input of 2676 words. The created network consists of 21 clusters, which encompass 155 terms and 622 co-occurrence links. Note that the study of the network allowed us to analyze patterns within the data and perform further data processing, including identifying, tagging, and removing generic and irrelevant terms. 

\begin{figure}
    \centering
    \hspace*{-1.0in}
    \includegraphics[width=1.29\linewidth]{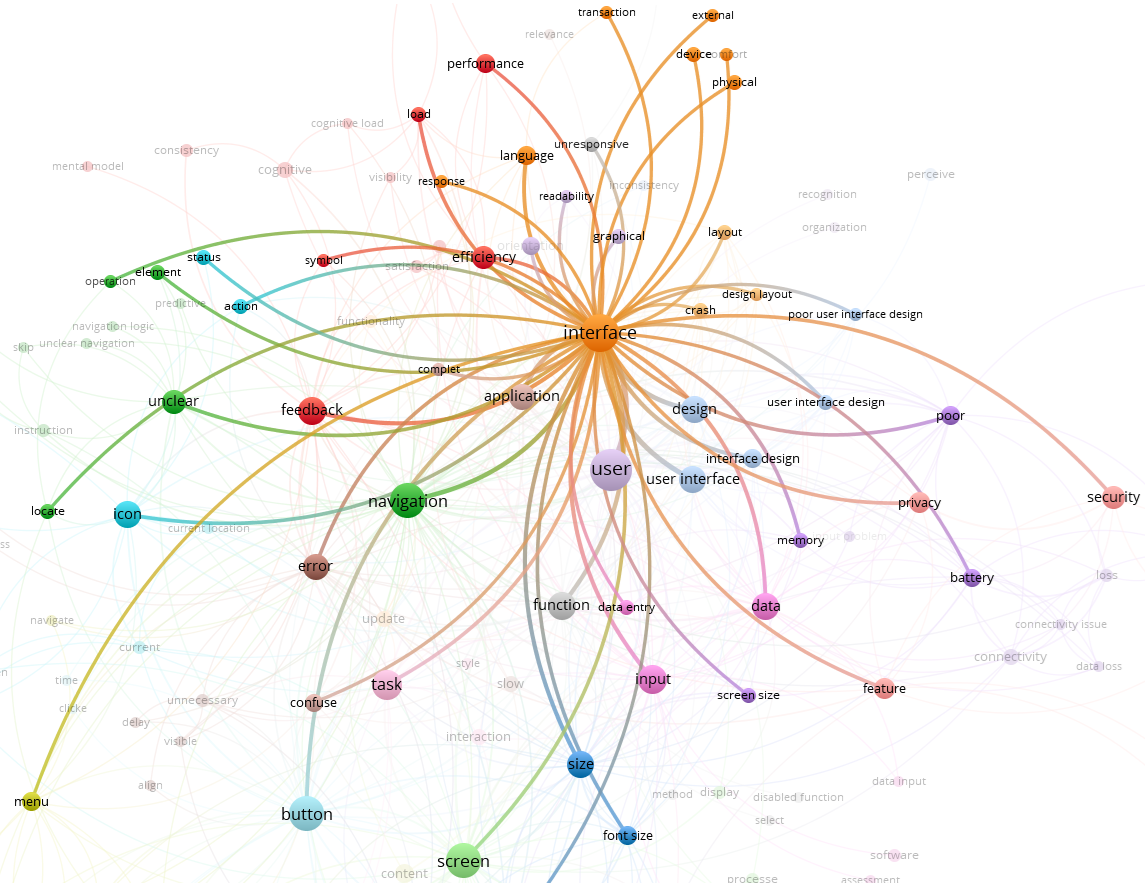}
    \caption{Visualization of the cluster for the term "Interface".}
    \label{fig:interface}
\end{figure}

\begin{figure}
    \centering
    \hspace*{-1.0in}
    \includegraphics[width=1.29\linewidth]{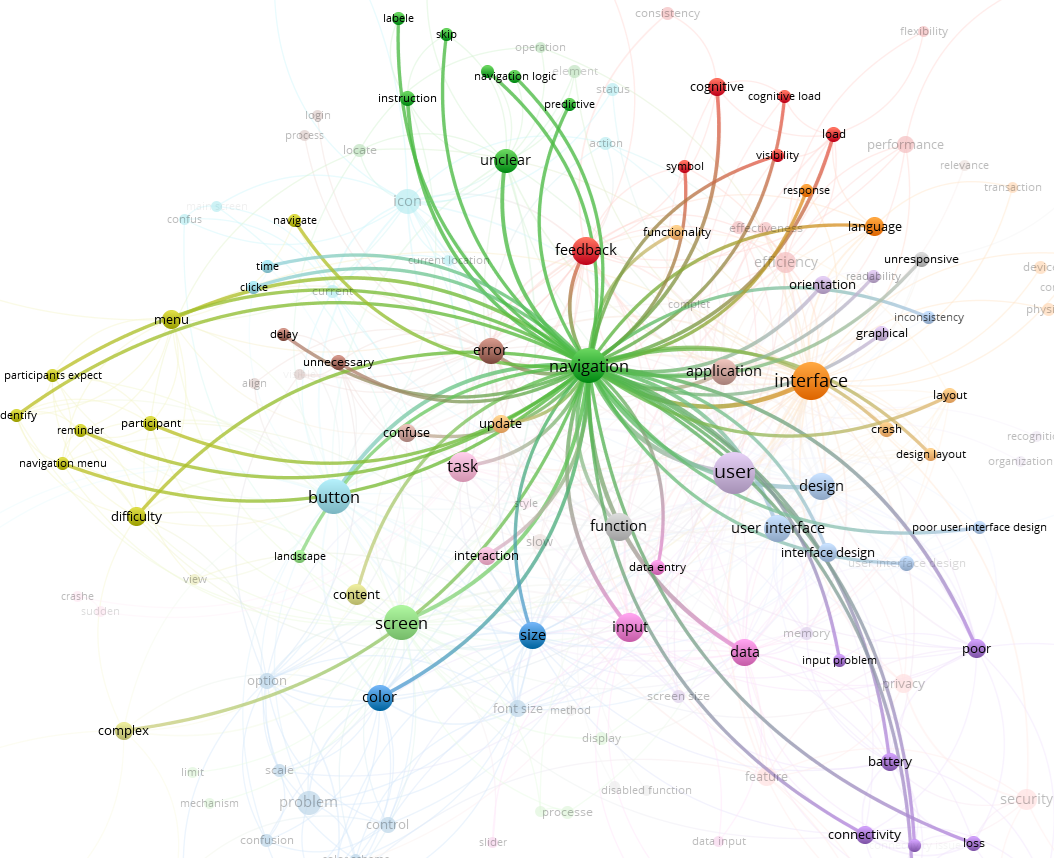}
    \caption{Visualization of the cluster for the term "Navigation".}
    \label{fig:navigation}
\end{figure}

We also recognize the value of the co-occurrence analysis which identifies relationships based on how frequently items appear together regardless of the strength or directionality of the connections. For instance, Figure~\ref{fig:interface} and Figure~\ref{fig:navigation} show the patterns for interface and navigation terms, respectively.

More specifically, Figure~\ref{fig:interface} shows Cluster 7, which has 55 links and 21 terms with a total link strength of 130. Here, the term "interface" is associated with terms that have a neutral meaning, such as application,  design, navigation, input, and feedback, as well as terms that have a negative meaning, such as unclear, error, confuse, poor, and crash, among others. 

Next, as shown in Figure~\ref{fig:navigation}, Cluster 2 contains 67 links and 17 terms, with a total link strength of 127.14. Similarly, the term navigation is associated with both neutral terms like user interface, task, screen, and feedback and negative terms such as difficulty, delay, crash, unclear, and unresponsive, to name a few.

It's important to note that exploring individual clusters within the network is an easy way to find relevant information. In general, we found that this software is a valuable research tool. However, further analysis is needed to gain detailed insights into the extracted dataset.

\subsection{Code Extraction}
In essence, in vivo coding assumes that codes are extracted directly from documents. This preserves the original language used in qualitative data. Such approach maintains authenticity by grounding the analysis in the reported lived experiences usability issues. By definition, a code is a phrase or expression of one or more words that encapsulates meanings relevant to the research question.

The following three steps were undertaken to extract codes. First, we carefully read the text line by line. In particular, we paid attention to phrases or expressions that stood out or were repeated. Second, we identified meaningful phrases by highlighting exact words or short phrases. Third, we transferred the codes to a spreadsheet to enable frequency counting and further analysis.

A total of 457 codes were extracted. An analysis of keyword density shows that the ten most frequent single words are: information (\#26), interface (\#24), user (\#22), screen (\#20), unclear (\#20), issues (\#18), navigation (\#17), lack (\#17), difficult (\#17), and application (\#15). The top five a two-word phrases are: user interface (\#11), difficult understand (\#10), interface design (\#5), font size (\#5), and main screen (\#4). 

The next step is semantic analysis, which involves interpreting the meaning of the extracted codes to identify general or specific usability issues. This process also bridges the gap between raw language and analytical interpretation to organize the codes into meaningful categories.

\subsection{Semantic Evaluation of Codes}
In qualitative research, semantic evaluation involves analyzing the meaning and relationships of the language used in a study. In our case, it is a method of identifying and interpreting the explicit meanings conveyed by the codes extracted from a dataset.
Note that for most qualitative studies, including those using in vivo coding, three to seven evaluation criteria are generally sufficient. Due to the nature of our research topic, we defined four exclusion and three inclusion criteria.

\subsubsection{Exclusion Criteria}
In the first pass, the following exclusion criteria guided the undertaken semantic evaluation:

\begin{enumerate}
    \item [EC1] Common or prevalent to the research topic: Exclude data that do not offer new or unique insights.
    
    \item [EC2] Lack of relevance to the research question: Exclude data that do not contribute meaningfully to answering the research question.
    
    \item [EC3] Context dependency or application specificity: Exclude data that are too narrowly tied to a specific context, application, or system, as this limits their generalizability.
    
    \item [EC4] User-specific actions or behavior. Exclude data that reflects highly individualized or idiosyncratic user action or behavior that cannot be reasonably generalized or replicated.
\end{enumerate}

The first exclusion criterion (EC1) applies to words or phrases commonly associated with the research topic. Specifically, codes tend to express well-known, generic, or obvious statements without providing additional detail, context, or perspective. For instance, the following single words were excluded: ``compatibility", ``device", ``flexibility", ``legibility", ``mobility", ``multimodality", ``issue", along with others, including ``emotional response", ``user confidence", and ``physical environment". 
In this case, 82 codes were removed from the input dataset.

The second exclusion criterion (EC2) pertains to irrelevant codes to the scope covered by the research question. In other words, these kinds of expressions fall within the scope of human-computer interaction; however, they do not meaningfully contribute to our study. For example, the following codes were excluded: ``comparison with other similar products",  ``dependency on the system", ``different display resolutions", ``hand manipulation", ``system vs. world", ``visual attention", among others. In this case, 25 codes were removed from the input dataset.

While checking the third exclusion criterion (EC3), interestingly, some authors documented only issues strictly related to the specific features of the mobile application tested by the study participants. In this vein, they did not attempt to generalize those issues into broader themes. For instance, ``complex, long sentences", ``does not have page name", ``game mechanics", ``grammatical error", ``lack of advance options", ``limited phone space for new messages", ``no KAI payment feature", ``quality of the image", ``redundancy wording", ``texts are not aligned", ``the text is not justified", ``words have been distorted".
These findings, represented by a total of 173 codes, were excluded based on the third criterion (EC3) due to their specific or individual nature.

Last but not least is the fourth exclusion criterion (EC4), which concerns codes that explicitly describe specific users' actions recorded during application testing, as well as their attitudes or perceptions expressed afterward. For instance, 
``cannot find category, so user typed in order to search", ``cannot save the favorite places",  
``confusion when needing to vertical scroll", ``did not see current location button", ``got lost into Measurement Tools function", ``participant unaware that tapping was an option", ``participants noted difficulty with dragging pictures in PSM", 
``recognition of the scale range is difficult", ``unable to show detail result page". In total, 28 codes were removed from the input dataset. 

In summary, applying all exclusion criteria resulted in the removal of 308 (67.40\%) codes. The input dataset now contains 149 codes, which will undergo further evaluation.

\subsubsection{Inclusion Criteria}
The inclusion criteria explicitly define the characteristics and conditions that a code must meet to be considered relevant and valuable for analysis. Note that, in this qualitative research, defining inclusion criteria (IC) involves transitioning from broad, theory-driven foundations to specific, context-driven conditions. At a general level, the first inclusion criterion (IC1) reflects alignment with the grounded theory in the field of Human-Computer Interaction (HCI). The second criterion (IC2) relates to the PACMAD+3 model at the model-driven level and encompasses ten usability attributes derived from cutting-edge literature on mobile HCI. The third criterion reflects alignment with general usability traits applicable to all mobile applications.
This layered approach balances openness to discovery with analytical focus and rigor.

Accordingly, we formulated the following inclusion criteria, which were subsequently verified during the second pass:

\begin{enumerate}
    \item [IC1] Alignment with Mobile Human-Computer Interaction Literature: Include any codes that correspond to recognized usability concepts, or themes discussed in the existing literature on mobile usability.
    
    \item [IC2] Adherence to the PACMAD+3 conceptual model: Include any codes that align with the model's theoretical usability attributes.
    
    \item [IC3] Direct reference to a generic usability trait: Include data that explicitly refers to widely recognized qualities within the scope of mobile application usability.
\end{enumerate}

In the first pass, we carefully examined each code by analyzing its linguistic semantics, taking into account the relevance to the research question. Thus, the following codes and words were removed: ``application" (\#4), ``lack" (\#2), ``problem" (\#2), user (\#1), complicated (\#1), ergonomics (\#1), cumbersome (\#1), features (\#1), ``issue" (\#1), ``use" (\#1), menu (\#1), required (\#1), product (\#1), ``software" (\#1), ``undo" (\#1). In addition, each code was pruned of adjectives with negative connotations that lacked substantial informative value such as ``poor" (\#4) and ``bad" (\#2). Next, we performed data stemming to simplify code dimensionality and enable their effective generalization. Afterwards, the dataset was examined against each inclusion criterion independently.

When applying the first inclusion criterion (IC1), we addressed the following question: Does the code reflect a concept or issue commonly addressed in mobile human-computer interaction (HCI) literature? A code that matches at least three studies published within the past ten years meets IC1. In this pass, all 35 codes met this condition. Next, we counted the frequency of each code. Seven unique themes emerged, each of which occurred at least twice, as shown in Table~\ref{tab:codes-IC1}.

\begin{table}[h]
\centering
\small
\caption{The list of themes related to mobile usability issues that met IC1.}
\label{tab:codes-IC1}
\begin{tabular}{|l|r|p{6.8cm}|}
\hline
\textbf{Theme}       & \textbf{Count} & \textbf{Reference} \\ \hline
Navigation                 & 12    &    \cite{kjeldskov2005evaluating},
\cite{wright2005usability}, \cite{nielsen2006s},
\cite{ehrler2018mobile},
\cite{alqahtani2019usability}, 
\cite{choemprayong2021exploring}, 
\cite{ziabari2021creating}, 
\cite{faudzi2022evaluating}, 
\cite{koowuttayakorn2022usability},
\cite{young2023remote},
\cite{wohlgemut2023methods},
\cite{donawa2024designing}
\\ \hline
Design                     & 8     & \cite{ehrler2018mobile}, \cite{alqahtani2019usability}, \cite{faudzi2022evaluating},
\cite{wohlgemut2023methods}, \cite{zhang2005challenges}, \cite{garcia2021experiment}, \cite{alqahtani2015investigation},
\cite{robson2020heuristic}
\\ \hline
Security                   & 5     &  \cite{robson2020heuristic}, \cite{billi2010unified}, \cite{attah2021mobile}, \cite{greene2003usability}, \cite{velez2014usability}      
\\ \hline
Privacy                    & 3     &  \cite{attah2021mobile}, \cite{greene2003usability}, \cite{o2014exploring}
\\ \hline
Help                       & 2     &  \cite{kjeldskov2005evaluating}, \cite{robson2020heuristic}
\\ \hline
Accessibility              & 2     &  \cite{ali2015study}, \cite{muller2022mobile}
\\ \hline
Information architecture & 2     & \cite{billi2010unified}, \cite{jeong2020detecting}           
\\ \hline
\end{tabular}
\end{table}

Considering the second inclusion criterion (IC2), we investigated the following question: Which specific attribute of the PACMAD+3 model does the code match? The attribute name and its conceptual definition serve as primary points of reference \cite{weichbroth2024usability}, along with well-recognized indicators used for its operationalization \cite{weichbroth2020usability}. In this case, 19 codes were identified. A total of eight attributes were recognized, although Memorability \cite{ismail2016review} and Satisfaction \cite{othman2018heuristic} appeared only once each. Table~\ref{tab:codes-IC2} shows the summary in this regard.

\begin{table}[h]
\centering
\small
\caption{The list of attributes related to mobile usability issues that met IC2.}
\label{tab:codes-IC2}
\begin{tabular}{|l|r|l|}
\hline
\textbf{Attribute}  & \textbf{Count} & \textbf{Reference} \\ \hline
Efficiency        & 5     &   \cite{kjeldskov2005evaluating}, \cite{katusiime2019review}, \cite{alsanousi2023investigating},
\cite{lynn2020increasing}, \cite{chin2021use}
\\ \hline
Ease of use       & 3     & \cite{zhang2005challenges}, \cite{baharuddin2013usability}, \cite{alhejji2022evaluating}
\\ \hline
Understandability & 3     &  \cite{ziabari2021creating}, \cite{alqahtani2015investigation}, \cite{wada2023feasibility}
\\ \hline
Cognitive load    & 2     &  \cite{kjeldskov2005evaluating}, \cite{nielsen2006s}
\\ \hline
Learnability      & 2     &  \cite{robson2020heuristic}, \cite{ali2015study}
\\ \hline
Effectiveness     & 2     &  \cite{katusiime2019review}, \cite{alsanousi2023investigating}
\\ \hline
\end{tabular}
\end{table}

Last but not least, considering the third inclusion criterion (IC3), we put forward the following question: Does the code clearly and directly refer to one of these recognized usability traits, either by name or description? We verified whether each code meaningfully aligned with the previously extracted mobile usability theme or attribute.
Next, we estimated the frequency of each attribute according to the number of studies. Note that the Navigation and Learnability categories are supported by one paper, while the others are covered by two or more sources. Table~\ref{tab:issues-grouped} shows more details in this regard. Ultimately, we determined and grouped 95 codes (usability traits) into 14 attributes, as presented in Table~\ref{tab:codes-extracted}. 

\begin{table}[h]
\centering
\small
\caption{The list of attributes related to mobile usability issues that met IC3.}
\label{tab:issues-grouped}
\begin{tabular}{|l|r|p{7.1cm}|}
\hline
\textbf{Attribute}  & \textbf{Count} & \textbf{Reference} \\ \hline
Design                   & 14       &   \cite{nielsen2006s}, \cite{ziabari2021creating}, \cite{koowuttayakorn2022usability} \cite{young2023remote}, \cite{ali2015study}, \cite{katusiime2019review},
\cite{othman2018heuristic}, \cite{caro2018identifying}, \cite{hussain2017usability},
\cite{shah2019systematic}, \cite{moradian2018usability}, \cite{lim2006comparative},
\cite{al2016heuristic}, \cite{al2019mobile}
\\ \hline
Efficiency               & 11       &   \cite{faudzi2022evaluating}, \cite{zhang2005challenges}, \cite{garcia2021experiment},
\cite{attah2021mobile}, \cite{chin2021use}, \cite{shah2019systematic}, \cite{lim2006comparative}, \cite{ahmad2021usability}, \cite{kim2015usability}, \cite{fergo2023evaluation}, \cite{chan2022user}
\\ \hline
Understandability        & 10       &    \cite{nielsen2006s}, \cite{ehrler2018mobile}, \cite{alqahtani2019usability}, \cite{imtinan2013usability}, \cite{thanachan2016comparative}, \cite{al2019mobile}, \cite{van2019usability}, \cite{meidani2024evaluating}, \cite{putra2018interaction}, \cite{yoo2016personalized}     
\\ \hline
Ease of use              & 9       &   \cite{alqahtani2015investigation}, \cite{attah2021mobile}, \cite{alsanousi2023investigating}, \cite{shah2019systematic}, \cite{van2019usability}, \cite{ahmad2021usability}, \cite{kim2015usability}, \cite{putra2018interaction}, \cite{bashir2019euhsa}
\\ \hline
Information architecture & 7    &  \cite{kjeldskov2005evaluating}, \cite{jeong2020detecting}, \cite{al2019mobile}, 
\cite{kim2015usability}, \cite{bashir2019euhsa}, \cite{monkman2013health}, \cite{schopfer2022mobile}
\\ \hline
Errors                   & 6        &    \cite{alqahtani2019usability}, \cite{alqahtani2015investigation}, \cite{velez2014usability}, \cite{alhejji2022evaluating}, \cite{shah2019systematic}, \cite{hsieh2019usability}
\\ \hline
Operability              & 6        &    \cite{robson2020heuristic}, \cite{shah2019systematic}, 
\cite{kim2015usability}, \cite{chan2022user}, \cite{schopfer2022mobile}, \cite{duh2006usability}
\\ \hline
Cognitive load           & 5        & \cite{nielsen2006s}, \cite{robson2020heuristic}, \cite{muller2022mobile}, \cite{lynn2020increasing}, \cite{heo2009framework}
\\ \hline
Effectiveness            & 4        & \cite{nielsen2006s}, \cite{alqahtani2015investigation},  \cite{baharuddin2013usability}, \cite{putra2018interaction}
\\ \hline
Accessibility            & 3        &    \cite{alsanousi2023investigating}, \cite{monkman2013health}, \cite{dunsmuir2019postoperative}
\\ \hline
Help                     & 2        &  \cite{alqahtani2019usability}, \cite{meidani2024evaluating}
\\ \hline
Memorability             & 2        & \cite{shah2019systematic}, \cite{lim2006comparative} 
\\ \hline
Learnability             & 1        & \cite{shah2019systematic}
\\ \hline
Navigation               & 1        & \cite{meidani2024evaluating}
\\ \hline
\end{tabular}
\end{table}

\begin{table}[]
\centering
\footnotesize
\caption{Codes extracted represented }
\label{tab:codes-extracted}
\begin{tabular}{|p{2.5cm}|r|p{12.5cm}|}
\hline
\textbf{Category}  & \textbf{\#} & \textbf{Code} \\ \hline
Design    & 20       & attractiveness; not consistent font; not consistent image; not   consistent heading; lack of information about privacy and security features;   user interface organization; visibility; unclear labels; too small font size;   small screen or font size; lack of effective color scheme; icon/symbol/graphical   representation issues; appearance / look of the product; unclear buttons; aesthetics;   poor interface design (annoying colors, lack of following standards in icon   design, small font size); text was too   small; unclear labeling of interface components; color patterns; font   inconsistency 
\\ \hline
Understandability        & 13       & feedback; icons ambiguity; confusion in understanding the terminology   or wording; information is not clear and is not enough; unclear meanings of   labels; terminology interpretation; unclear control and confirmation; application's   interface is inconsistent and confusing; unclear instructions; language   issues; dialogue boxes labeled in an unclear way; buttons labeled in an unclear way; user interface misunderstanding                                   \\ \hline
Efficiency               & 12       & performance (\#3); slow responsiveness of the application; efficiency   to input data; information processing delays; slow loading times; efficient   interactions; slow response; long loading process; delayed notification   delivery; freezing of the app during use                                                                           \\ \hline
Ease of use              & 10       & difficulties in performing tasks (some of them were not agile nor   easy to complete); complicated menu; application very cumbersome to use; not   easy access to all features; visibility (not being able to find a particular   link, button, piece of information or a particular page); dense walls of   text; simplicity; consistency; difficult to understand application's   language; no clues how to perform a task                                                            \\ \hline
Information architecture & 8        & readability; the amount of GUI information displayed; information   localization and relevance; lack of topic organization; effective information   presentation; unnecessary information; grouping or distinction of items; information density                                                                                                       \\ \hline
Errors     & 7        & bugs: crashing forcefully closing; functionality errors; syntactic   errors; frequent crashing/software malfunction; deadlocks; correctness   (violation of syntax, unresponsive functions, links and buttons, or the   application failing to do what it is supposed to do and crashing); no visible   feedback                                                      \\ \hline
Operability   & 6        & contextual awareness; buttons not working properly; feedback;   preservation of context; unresponsive touch screens; immediate feedback                                                                                                 \\ \hline
Cognitive Load           & 5        & consistency (\#2); information overload; extensive workload in task   performance; highly complex tasks to perform                                                                                                          \\ \hline
Effectiveness            & 4        & lack of task support; task flow; completeness (lack of required   information); no feedback from application after inputting any action                                                                                       \\ \hline
Accessibility            & 3        & voice recognition; color contrast issues; trouble hearing alerts                            \\ \hline
Help                     & 2        & lack of guidance and explanation; poor assistance and support                               \\ \hline
Learnability             & 2        & items with overlapping concepts; difficulty in learning to use the   application             \\ \hline
Memorability             & 2        & complex hierarchal presentation and too many options to choose from; locating   appropriate interface elements                                                                                                                 \\ \hline
Navigation               & 1        & unclear navigation logic                                                                    \\ \hline
\end{tabular}
\end{table}

Further analysis and interpretation are discussed in the next section.

\subsection{Findings}
\label{subsec:study1-findings}
At this point, let us recall the research question, which was formulated as follows: What are the usability issues associated with mobile applications? Since there is no single answer to this question, the findings have been semantically grouped and discussed in separate subsections for greater clarity and readability.

\subsubsection{Design}
In the User Interface Design category, issues include inconsistent visual elements such as fonts, images, and headings, as well as an unappealing interface, unclear or malfunctioning buttons, and difficult-to-navigate menus. It was also noted that the design lacks visual coherence and that the text is sometimes too small or illegible. Other issues include ineffective color patterns and schemes and general design clutter.

\subsubsection{Understandability}
The identified issues within the Understandability and Help categories highlight significant challenges related to the clarity and interpretability of mobile application interfaces. Users often encounter vague terminology and unclear instructions. They also experience a lack of feedback, incomplete or poorly labeled information, and confusing requests. 

\subsubsection{Efficiency}
The issues identified within the Efficiency category hinder the performance of mobile applications, affecting their smooth and prompt operation. Users report slow response times, delayed content loading, inefficient data input, and interruptions, such as apps freezing or delayed notifications.

\subsubsection{Ease of use}
The Ease of Use category concerns a user perception on how intuitively and effortlessly can interact with a mobile application. Reported issues insufficient helpful cues, and inadequate support while using the app.

\subsubsection{Information architecture}
The information architecture category covers how content within a mobile application is structured, presented, and made accessible to users. Major usability issues in this area include overwhelming information density, poorly grouped related content, and unclear localization or placement of important information.

\subsubsection{Errors}
The Errors category includes technical and functional issues that disrupt the normal operation of a mobile application. Users  experience crashes, deadlocks, unresponsive functions, or incorrect system behavior.

\subsubsection{Operability}
Operability in mobile applications refers to the degree to which users can effectively operate and control it. Issues in this area include unresponsive or malfunctioning touchscreen buttons, inconsistent or absent feedback mechanisms. In addition, users also often report a lack of contextual awareness or failure to preserve context.

\subsubsection{Cognitive load}
In the context of mobile application usability, cognitive load refers to the mental effort users expend to process information, navigate tasks, and make decisions while interacting with the application. Users often struggle to complete tasks due to an overwhelming amount of information, complex tasks, or unfamiliar content.

\subsubsection{Effectiveness}
The identified issues in the Effectiveness category include unresponsive controls, poor task flow, lack of feedback, and difficulty locating key user interface elements.

\subsubsection{Accessibility}
Issues in the Accessibility category generally highlight common barriers and inadequate accommodations for a wide range of user needs. More specifically, these issues include insufficient voice recognition support, poor color contrast, and difficulty hearing alerts and notifications.

\subsubsection{Help}
In terms of available Help, users reported lack of guidance and explanation, and poor assistance or support.

\subsubsection{Learnability}
In terms of learnability, users report issues such as overlapping concepts, unfamiliar symbols, and inconsistent designs when performing a single task. 

\subsubsection{Memorability}
In regard to memorability, users have difficulty with overly complex hierarchies and an abundance of options. They also have trouble locating interface elements and recalling experiences with similar yet different applications.

\subsubsection{Navigation}
Issues in the Navigation category include unclear navigation.

\subsubsection{Security}
The issues identified within the Security category raise common concerns about how mobile applications interact with device-level protections and manage user data security. Users are primarily concerned about the lack of information security mechanisms and transparency regarding how their personal data is protected. Other concerns include infrequent updates that leave vulnerabilities unaddressed and policies that enforce insecure practices, such as mandatory password changes.

\subsubsection{Privacy}
The issues identified within the Privacy category emphasize the need for greater transparency and user awareness of data protection practices in mobile applications. More notably, the absence of easily accessible privacy policies and terms of use prevents users from understanding how their personal information is collected, stored, and shared.

\subsection{Summary}
Note that not all of these 16 categories are widely recognized in the field of mobile human-computer interaction. 
However, we will leave the original names unchanged since our goal, at this point, is neither to synthesize nor to introduce new categories. Besides, We deliberately did not bring up satisfaction, which is an individual feeling of contentment, happiness, or fulfillment that comes from achieving a desire, fulfilling a need, or resolving a problem. To sum up, these 16 categories will serve as the baseline for further analysis. 

\subsection{Other Findings}
In addition, two other group of factors, not directly related to mobile applications, emerged during the review. In case of the former, Table~\ref{tab:external-factors} presents seven external factors that have been reported to significantly impact the usability of mobile applications.

\begin{table}[]
\caption{List of external factors affecting usability of mobile applications}
\label{tab:external-factors}
\begin{tabular}{|l|l|l|}
\hline
\textbf{Factor} & \textbf{Count} & \textbf{References} \\ \hline
Low network speed   & 12    &  \cite{alqahtani2019usability, faudzi2022evaluating, garcia2021experiment, alqahtani2015investigation, billi2010unified, katusiime2019review, caro2018identifying, ahmad2021usability, kim2015usability, duh2006usability,al2015systematic,kukulska2007mobile} 
\\ \hline
Short battery life  & 8  & \cite{alqahtani2019usability,billi2010unified,velez2014usability,alhejji2022evaluating,caro2018identifying,al2016heuristic,imtinan2013usability,chan2022user}
\\ \hline
Screen size  & 6  &   \cite{billi2010unified,ismail2016review,alhejji2022evaluating,ahmad2021usability,al2015systematic,gafni2009usability}         
\\ \hline
Memory              & 6     &  \cite{alqahtani2019usability,heo2009framework, gafni2009usability,velez2014usability,imtinan2013usability,alhejji2022evaluating}        
\\ \hline
Processing power    & 4     &  \cite{heo2009framework,gafni2009usability,ismail2016review,katusiime2019review}, \\ \hline
Size of the physical buttons & 2     & \cite{alqahtani2019usability,shah2019systematic}   \\ \hline
Unstable connection & 2     &  \cite{faudzi2022evaluating, kim2015usability}          \\ \hline
\end{tabular}
\end{table}

Analyzing the demonstrated results, the following conclusions can be drawn. Fast network connectivity is undeniably a must-have feature for modern mobile devices to ensure a satisfying user experience. Slow data transfer, on the other hand, is the most significant external factor affecting the usability of mobile applications. As one might expect, the other factors concern limitations inherent to mobile hardware devices, including battery life, screen size, memory, and CPU capacity, along with the size of the physical buttons available for a user to control an application.

The second group concerns factors that reflect user perceptions experienced through interaction. In this space, one can point to: 
\begin{itemize}
    \item positive emotions such as: confidence \cite{kjeldskov2005evaluating}, comfort \cite{lim2006comparative}, trust \cite{hussain2017usability}, satisfaction \cite{ismail2016review}, and
    \item negative emotions such as: frustration \cite{young2023remote}.
\end{itemize}

In numerous studies, users' emotions have been conceptualized and tested as the outcomes (independent variables), underlying popular technology acceptance models \cite{lu2019exploring}.

\newpage
\section{Study \#2. Interviews with practitioners}
\label{sec:study2-interview}
Keeping in mind the goal of our study, the second phase of our research involved interviewing practitioners from the mobile software industry. Although our interest remained the same, we aimed to collect primary data on post-development lessons related to usability issues reported by users of released mobile applications. 
That being said, we put forward the following research question (RQ):

\begin{itemize}
    \item[RQ2:] What are the most common usability issues in mobile applications?
\end{itemize}

It should be noted here that to maximize the generalizability of qualitative research findings to broader contexts, we intentionally used the \textit{most common} words to abstract from specific issues related to particular application features.


\subsection{Research Method}
Considering the qualitative nature of the research topic, we used an interview as a qualitative research method with open-ended and predetermined questions to collect information from respondents. We chose open-ended questions to provide the necessary unbounded space to identify usability issues and maintain the scope and depth of our research data. In other words, we did not demand one correct answer; instead, we accepted any answer \cite{husain2012construct}. However, this type of question requires expertise and diligence from the interviewee. Therefore, respondents must exhibit a certain level of knowledge and experience. To reach such people, the obvious choice was to use purposive sampling. We believe that this data collection method and its settings are both valid, as such approach has the potential to yield valuable responses.

\subsection{Data sampling and time frames}
By definition, purposive sampling strategy ensures that specific kinds of cases of those that could possibly be included are part of the final sample in the research study \cite{campbell2020purposive}. In this line of thinking, we employed this strategy based on the assumption that different groups of people have valuable, unique perspectives on the usability issues under consideration and should therefore be included in the sample \cite{robinson2014sampling}. 
From September to December 2024, we reached potential interviewees through our personal contacts as well as social media.

\subsection{Sample characteristics}
To determine and establish a group of experts, we qualified interviewees based on their professional experience in the mobile software industry and the number of projects they had worked on in mobile application development. In this view, an expert is someone who has at least two years of experience in two different projects. From a pool of 20 people who were contacted and agreed to participate, we positively classified nine (four men and five women) for the survey. 
Table~\ref{tab:expert-sample} shows the detailed characteristics of the assembled pool of experts.

\begin{table}[h]
\centering
\small
\caption{A summary of the demographic information and professional backgrounds of the interviewed experts.}
\label{tab:expert-sample}
\begin{tabular}{|l|l|l|l|l|l|l|}
\hline
\textbf{Id} & \textbf{Gender} & \textbf{Age} & \textbf{Education} & \textbf{Current Occupation}    & \textbf{\#EX} & \textbf{\#P} \\ \hline
E1        & Woman & 41  & Higher    & Technical Writer          & 16        & 2        \\ \hline
E2        & Woman & 42  & Higher    & UX Writer                 & 16        & 2        \\ \hline
E3        & Man   & 26  & Higher    & Product Designer          & 5         & 2        \\ \hline
E4        & Man   & 42  & Hihger    & CPO                       & 10        & 2        \\ \hline
E5        & Woman & 26  & Hihger    & UX Designer               & 3         & 5        \\ \hline
E6        & Woman & 41  & Hihger    & UX Writer                 & 20        & 3        \\ \hline
E7        & Woman & 26  & Hihger    & UX Researcher             & 4         & 3        \\ \hline
E8        & Man   & 37  & Higher    & IT Consultant             & 20        & 5        \\ \hline
E9        & Man   & 22  & Higher    & Mobile Software Developer & 2         & 2        \\ \hline
\end{tabular}
\end{table}

The average age of the respondents was 33.7 years, and all of them had received a higher education. They currently hold a variety of positions ranging from general to highly specialized. These include consultants, designers, developers, and writers working in the software industry, as well as a researcher working at a technical university. 

Note that none of the experts agreed to disclose any personal or company information, including product or project names or budget sizes. Nevertheless, it is important to note that our respondents neither share a workplace nor collaborate on past projects.

\subsection{Data Collection}
To make things easier and more comfortable for everyone, we sent out an electronic questionnaire by email to each expert individually. 
The survey had three parts. The first part was the introduction and contains the guiding theme.
The second part asked respondents to answer the following question: What are the most common usability issues in mobile applications? The third part asked for demographic information, including age, level of education, current occupation, and years of experience. Additionally, participants were requested to identify and describe IT projects they have been involved in over the past three years.

Due to the clear and specific objective of our study, along with specialized group of respondents, we did not plan for follow-up discussions or evaluations. However, the experts were instructed to provide any important, detailed information to the parties involved in mobile software development and maintenance.

The experts provided feedback in both Polish and English. For the former language, we used two online American English translators, namely: Google Translate (translate.google.com) and DeepL (www.deepl.com/translator). Each sentence was independently translated by two translators. The results were then compared to check and ensure consistent terminology. Overall, the quality was highly satisfactory, with only a few translations requiring minor revisions.
Table~\ref{tab:interview-data-set} summarizes the collected data in quantitative terms, showing the estimated number of submitted sentences and words. Note that bullet points are not counted as full standalone sentences.

\begin{table}[h]
\centering
\small
\caption{A summary of the collected responses by gender, including the number of sentences and words.}
\label{tab:interview-data-set}
\begin{tabular}{|l|l|r|r|}
\hline
\textbf{Id} & \textbf{Gender} & \textbf{\#Sentences} & \textbf{\#Words} \\ \hline
E1 & Woman  & 3         & 19    \\ \hline
E2 & Woman  & 5         & 40    \\ \hline
E3 & Man    & 10        & 270   \\ \hline
E4 & Man    & 3         & 20    \\ \hline
E5 & Woman  & 5         & 21    \\ \hline
E6 & Woman  & 6         & 61    \\ \hline
E7 & Woman  & 4         & 67    \\ \hline
E8 & Man    & 19        & 299   \\ \hline
E9 & Man    & 6         & 60    \\ \hline
\end{tabular}
\end{table}

Our sample includes 857 words and 61 sentences. The median of answer length, measured by word count, is 60. Among women, the lengths appear consistent, whereas among men, two outliers (E3 and E4) are noticeable. 
Similarly, from a qualitative perspective, the collected data revealed significant variation in both the style of expression and the level of detail. In other words, while some respondents provided brief, bullet-pointed answers, the others opted for a more detailed, narrative style.

\subsection{Data Analysis}
First, we carefully reviewed all of the collected data. Preliminary screening revealed that all responses were logical, evidence-based, and pragmatic, proving their relevance and validity.

To address the research question in an appropriate manner, we used an in-vivo coding (also termed as literal coding) approach. The rationale for its selection is based on the following premises. First, in-vivo coding is a qualitative data analysis method in which the exact words or phrases of the participants are used as codes \cite{chakraborty2023we}. This approach preserves the authenticity and voice of respondents, making it especially valuable in exploratory research \cite{gupta2024codes}. Third, in vivo coding is often used to develop grounded theory \cite{manning2017vivo}. Therefore, in-vivo coding was a suitable and reasonable choice for extracting mobile usability issues.

In-vivo coding was performed as follows \cite{rivas2012coding,ribeiro2014using}:
\begin{enumerate}
    \item Reading: A careful reading to understand the context, paying attention to recurring ideas, emotions, and striking phrases.
    \item Coding: Extract individual words, topics, or themes and label them with double quotation marks as codes.
    \item Evaluating and pruning: Evaluate the relevance of each code and remove the irrelevant ones from the dataset.
    \item Counting: Determine the number of extracted codes.
    \item Grouping: Analyze the meaning of each code and group similar in vivo codes together to begin forming broader attributes.
\end{enumerate}

From a practical standpoint, the first two steps are performed simultaneously. Additionally, we printed the spreadsheet, which allows for highlighting and adding comments. When examining individual whole sentences or single bullet points, we separated individual issues or single topics using double quotation marks. 
In the first pass, a total of 99 codes were extracted. In the second pass, we evaluated each set of codes extracted from a particular response. We considered the context of the entire sentence, as well as the merit and relevance of each code. This process resulted in the removal of 26 codes, leaving 73 in the data volume.

It should be noted that Expert\#3 identified the intentional use of unethical practices by the software provider as the primary obstacle specifically targeting less experienced users. While we strongly disapprove of such practices, we excluded these codes from the analyzed dataset, assuming they are isolated incidents.

In-vivo coding involves grouping similar or related codes derived from participants' exact words or phrases, helping to organize the data and identify patterns across responses. In our study, these groups are structured according to usability attributes or, more broadly, by categories that reflect shared concerns and issues. Since the grouping process closely aligns with thematic analysis, we draw on the previously identified mobile usability attributes and categories (see Section~\ref{subsec:study1-findings}). For greater clarity and coherence, we will present the details of the further analysis and the classification results in the next section.

\subsection{Results}
First, we performed a semantic analysis to identify shared meanings between the codes and categories. Next, each code was then assigned to a unique attribute where a clear match was found. 

During analysis, four other attributes emerged. The first is Language, which encompasses seven codes: grammar errors, punctuation errors, stylistic errors, texts that are too formal, texts that are too official, texts that do not match the product's voice and tone, and non-inclusive texts. However, considering the categories that have already been identified, these codes fall within the scope of the Understandability category. 

The second is Responsiveness which contains four codes: unresponsive elements, inadequate adaptation to mobile screens, lack of responsiveness, and lack of optimization for mobile devices. 
Since responsiveness refers to a mobile application user's ability to complete assigned tasks within a given time, this category falls within the scope of Effectiveness.

The remaining two: Advertising, and Invasiveness, each have one code, respectively: intrusive ads, and too many notifications. While the former refers to a commercial and public medium promoting a specific product or service, it is an element of user interface design, however do not imply to the mobile applications in general. Similarly, notifications are a mechanism that sends alerts and messages to a user's device to provide updates or timely information. Nevertheless, they still remain a convenient option for certain applications. Thus these two codes will not be further considered.

Table~\ref{tab:study2-cat-codes} shows the results of the thematic analysis, based on these assumptions. Eventually, ten categories emerged from the analysis of the experts' responses.

\begin{table}[H]
\footnotesize
\caption{Mobile usability attributes, sorted in descending order by the number of assigned codes.}
\label{tab:study2-cat-codes}
\begin{tabular}{|p{2.7cm}|l|p{10.5cm}|}
\hline
\textbf{Category} & \textbf{Count} & \textbf{Codes} \\ \hline
Design & 22 & 
unnecessary animations; inconsistency between the mobile and web versions; use of dark patterns; fonts that are too small; failure to comply with minimum contrast levels; aesthetic considerations over accessibility; lack of consistency with generally accepted standards; large graphics and headers; cluttered screens; inconsistent texts; inconsistent image of the application; mobile devices limit layout options; need for good visual hierarchy; opaque UI; opacity due to lack of affordances; skeuomorphic affordances; chiseled 3d button; modern interface hides scrollbars; buttons indicated by body text; placing clickable objects close to the edge of the screen; displaying tables on small screens; unintuitive layouts 
\\ \hline
Understandability & 13 & interface element naming; abbreviations without explanation; icons without explanation; texts that are too technical; texts that are too enigmatic; special knowledge required to understand expression; grammar errors, punctuation errors; stylistic errors; texts that are too formal; texts that are too official; texts that do not match the product's voice and tone; non-inclusive texts
\\ \hline
Effectiveness & 8 & not enough focus on core actions; not supporting users in the process; user is unaware that a process has started or is in progress; unintuitive processes; unresponsive elements; inadequate adaptation to mobile screens; lack of responsiveness; lack of optimization for mobile devices 
\\ \hline
Operability & 7 & designed with desktop in mind; long scrolling; buttons that do not work 
\\ \hline
Information architecture & 6 & inconsistent product sorting; not enough focus on core features; unclear information hierarchy; need to prioritize content; no preloaders or progress bars; lack of appropriate error messages 
\\ \hline
Errors & 6 & app crashes; login errors; logging out of the application for no reason; errors during password recovery; bugs; poor error handling \\ 
\hline
Efficiency & 4 & poor loading time; long application opening time; long loading times for individual subpages; freezing \\ \hline
Accessibility & 2 & failure to comply with digital accessibility guidelines; failure to comply with WCAG \\ \hline
Ease of use & 1 & poor navigation \\ \hline
Learnability & 1 & users have to learn from scratch \\ \hline
\end{tabular}
\end{table}

The Design attribute includes the most codes, with a total of 22. The next categories are Understandability (\#13), Effectiveness (\#8), Operability (\#7), Information architecture and Errors (each \#6), Efficiency (\#4), and Accessibility (\#2). In the end, Ease of use and Learnability contain one code each. As one can easily conclude, these ten attributes only encompass 71 of the 73 codes. 

\section{Synthesis and Generalization of Results}
\label{sec:synthesis-generalization}
In the previous sections, using a grounded theory approach, we identified and analyzed 16 distinct categories of usability issues based on two data sources.
Since our goal is to develop a generic and unified a catalog of usability issues while avoiding unnecessary complexity, thus, one has to test and establish both internal and external validity.

\subsection{Internal Validity}
By definition, internal validity corresponds to credibility and is described as the best available representation of accuracy~\cite{arslan2025validity}. In other words, internal validity concerns the extent to which research findings correspond with reality and reflect what actually exists~\cite{bryman2016social}.
To determine internal validity, we examine the collected evidence from two performed studies. 
To this end, we used a two-stage approach. 

In the first stage, we perform a conceptual inter-analysis by critically comparing existing conceptualizations with the identified categories and their corresponding keyword dictionaries.
In the second stage, we conduct a conceptual intra-analysis by identifying underlying relationships and patterns within the extracted categories. Specifically, we analyze each category and compare it with the others, drawing on both web sources and scholarly literature. The results of the semantic analysis, comparison and merging are as follows:
\begin{itemize}
    \item User Interface Design and Information Architecture to \textbf{User Interface Design}. According to Wikipedia \cite{wikipedia-2025}, ''(...)~user interface (UI) design primarily focuses on information architecture. It is the process of building interfaces that clearly communicate to the user what's important". Similarly, recent research has also advocated this view~\cite{saparamadu2021user,farooq2025improving}.

    \item Understandability with Help to \textbf{Understandability}. These two categories share a significant amount of meaning. The former relates to the perceived value of information that aids the user in performing a task. Modern mobile applications, on the other hand, generally lack a traditional "Help" section~\cite{muller2014won}, which is common in desktop software. Instead, mobile applications offer direct assistance through features embedded in the user interface such as contextual hints~\cite{yan2023muid}, guided walkthroughs~\cite{fuller2021mobile} or progress bars~\cite{li2024impact}. 
    
    \item Operability with Navigation to \textbf{Operability}. The former is defined as a mobile application's ability to be operated by users~\cite{weichbroth2024usability}, while the latter refers to the way users navigate an application on their mobile devices. Both terms relate to the user interface and the physical buttons available on the device. However, operability is an already well-established attribute of mobile applications~\cite{weichbroth2024usability}. In this vein, some studies have introduced suitable heuristics under the operability measure~\cite{hashim12025mobile}, while others tested and evaluated the operability~\cite{arnhold2014mobile}.
\end{itemize}

To sum up, from the initial six categories, only three, namely: User Interface Design, Understandability, and Operability, will be further considered. 
Since the remaining categories are well-established in the scholarly literature, one can conclude that internal validity has been achieved. According to Khorsand and Crawford~\cite{khorsan2014external}, internal validity is a prerequisite for external validity. Thus, we can proceed to examine the latter.

\subsection{External Validity}
By definition, external validity is defined as "the extent to which the results of a study can be generalized beyond the specific context of the study to other populations"~\cite{gelo2008quantitative}. 
In our study, external validity concerns the applicability of the results to the general model of mobile application functionality.

According to the recent studies of Harrison et al.~\cite{harrison2013usability} and Weichbroth~\cite{weichbroth2024usability}, well-recognized in the current stream of research, it is clear that the Accessibility, Security, and Privacy categories do not meet the conditions of generalizability and applicability. Furthermore, the typical mobile app user is not equipped with both tools and knowledge necessary to evaluate the privacy and security of mobile applications. 
Therefore, we will not consider these three categories any longer.

\section{Results}
\label{sec:final-results}
The current study aimed to answer the following question: What usability issues are associated with mobile applications? 
Drawing on a triangulation approach that incorporates two data sources and research methods, below we introduces a catalog of the usability issues in modern mobile application development, classified under a three-tier system. In addition, we also present a novel taxonomy, a controlled vocabulary organized into a hierarchical structure consisting of usability issue categories and their corresponding keywords.

\subsection{Catalog and classification of usability issues}
\begin{figure}[H]
    \centering
    \includegraphics[width=.9\linewidth]{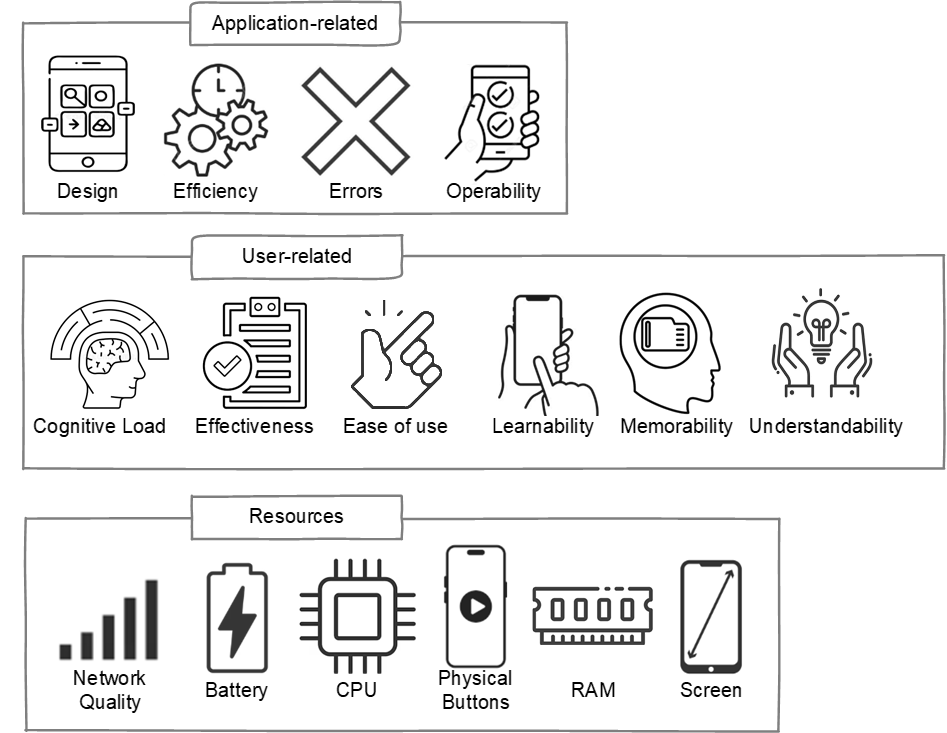}
    \caption{The three-tier AUR classification and the catalog of usability issues in modern application development.}
    \label{fig:classification}
\end{figure}

In summary, of the sixteen input categories, ten remained. Note that these categories correspond to the general attributes of the mobile application. 
Figure~\ref{fig:classification} presents the three-tier Application–User–Resource (AUR) classification system for the identified mobile usability issues.

The top tier is related to the four attributes of mobile applications. 
User interface design encompasses the layout and structure of information, the component arrangement, and visual styling. 
Efficiency refers to an application's ability to respond quickly and accurately to user actions. 
Errors refer to the number of defects and malfunctions that occur during a user's performance of a task. 
Operability corresponds to context sensitivity, which reflects an app's ability to adapt to and be used in different circumstances. 

The middle tier is associated with the six user-oriented abilities.
Effectiveness is defined as a user’s ability to successfully complete a task within a specific context.
Cognitive load is the amount of mental effort required by a user's working memory when using a mobile application. 
Ease of use refers to the extent to which a user perceives that using a mobile application requires no unnecessary effort.
Learnability refers to how easily a user can engage with a new app without external guidance or help.
Memorability is related to a user’s ability to effectively remember how to use an application.
Understandability is the ability to easily comprehend any part of a mobile application.

The bottom tier concerns available resources. 
First and foremost, many applications rely heavily on internet data, which significantly influences their availability and performance.
Battery drain limits how long mobile applications can be used.  
High CPU usage can be caused by demanding tasks within a single application, a software glitch, or multiple applications running simultaneously or in the background. This slows down the system and negatively affects the efficiency of running applications. Similarly, usability issues may also occur in the event of a shortage of RAM memory.
Another issue is that the physical buttons are not always accessible due to their small size.
Lastly, the size of the mobile device screen is also perceived as a significant limitation when operating a mobile application.

Additionally, a smartphone or other hardware device is usually equipped with an operating system, such as Android or iOS. Therefore, from a user's point of view, this factor can also be considered an issue.


\subsection{Taxonomy}
To organize and categorize information within a domain of mobile usability issues, we also decided to develop a taxonomy, based on the aforementioned classification system. Specifically, considering the nature of the available data, our taxonomy brings the form of a controlled vocabulary. Since a controlled vocabulary is a carefully curated and standardized set of terms, we identified and removed all context-sensitive items during the extraction and integration. 

At its most basic, a controlled vocabulary is used to standardize communication and facilitate understanding. In our study, the aim of a taxonomy is to identify and select appropriate terms to describe the content and nature of a resource, thereby supporting the discoverability and resolvability of the issue.

To achieve this, we need to determine and collect all discrete terms relevant to each mobile usability category. To this end, we combined the data from Table \ref{tab:codes-extracted} and Table~\ref{tab:study2-cat-codes}. In this way, 228 keywords were extracted, which are presented in Table~\ref{tab:taxonomy}.

\begin{table}[]
\centering
\footnotesize
\caption{Taxonomy of mobile usability issues.}
\label{tab:taxonomy}
\begin{tabular}{|l|l|p{11cm}|}
\hline
\textbf{Group}    & \textbf{Category}  & \textbf{Keywords}  \\ \hline
\multirow{4}{*}{\rotatebox[origin=c]{90}{\parbox[c]{7cm}{\centering Application}}}
& Design  & architecture, bar, color, component, consistent, contrast, dark,   dense, density, design, display, feature, font, form, graphic, grouping,   header, heading, hierarchy, icon, image, information, interface, interface,   item, label, labeling, layout, localization, look, message, object,   organization, organization, pattern, prioritize, progress, readability,   relevance, screen, scrollbar, size, small, sort, spelling, standard, symbol,   text, topic, unclear, user, visibility \\ \cline{2-3} 
 & Efficiency        & closing, data, delay, efficiency, fast, freezing, input, loading,   long, notification, opening, performance, poor, processing, response,   responsive, slow, speed            \\ \cline{2-3} 
  & Errors  & bug, crash, deadlock, error, fix, function, handling, incorrect,   malfunction, recovery, syntax, terminate, violation                                                \\ \cline{2-3} 
 & Operability   & adapt, aware, battery, button, clickable, context, control, control,   feedback, gesture, gesture, haptic, interactive, localization, navigate,   network, operation, optimized, orientation, responsive, screen, scrolling,   swipe, touch, unresponsive                 \\ \hline
\multirow{6}{*}{\rotatebox[origin=c]{90}{\parbox[c]{8cm}{\centering User}}}
& Cognitive Load    & capacity, clarity, cognition, complex, consistent, extensive,   hierarchy, intuitive, mental, overload, simplicity, strain, think, unnatural,   workload                                                                                              \\ \cline{2-3} 
  & Effectiveness     & action, adapt, assistance, attainment, completion, complex, correct,   documentation, effective, effectiveness, feedback, flow, goal, input,   instruction, instruction, interaction, optimization, outcome, perform,   process, tip, troubleshooting, unintuitive, voice   \\ \cline{2-3} 
 & Ease of use       & complete, consistency, discoverability, drawer, feature, feedback,   flow, home, locate, logic, map, menu, path, perform, search, status,   structure, task                    \\ \cline{2-3} 
 & Learnability      & affordance, familiar, faq, habit, learn, onboarding, overlapping,   pattern    \\ \cline{2-3}  & Memorability      & memory, progressive, recall, recognition, recognize, recurring,   remember, retention                                                                    \\ \cline{2-3} 
& Understandability & abbreviation, ambiguity, box, button, confirmation, confusion,   contact, control, dialogue, form, formal, grammar, guide, help, icon,   interpretation, label, language, misunderstanding, naming, official, punctuation,   reaction, stylistic, support, terminology, text, tone, tutorial, typos,   understanding, walkthrough, wording                                          \\ \hline
\multirow{6}{*}{\rotatebox[origin=c]{90}{\parbox[c]{1cm}{\centering Resource}}}
& Network Quality   & connection, network, speed  \\ \cline{2-3} 
  & Battery           & battery                                                                      \\ \cline{2-3} 
 & CPU               & power, processor, processing \\ \cline{2-3} 
 & Physical Buttons  & button, physical                                                              \\ \cline{2-3} 
 & RAM   & operating, memory \\ \cline{2-3} 
 & Screen  & touch, screen, size \\ \hline
\end{tabular}
\end{table}

In summary, Table~\ref{tab:taxonomy} provides a unified arrangement of mobile usability categories and related keywords in taxonomic groups according to their extracted relationships.
The developed taxonomy illustrates the terminology used to describe the sixteen categories of usability issues related to mobile applications.

\section{Discussion}
\label{sec:discussion}
\subsection{Theoretical implications}
In light of recent advancements in mobile human–computer interaction, particularly the PACMAD+3 model introduced by Weichbroth~\cite{weichbroth2024usability}, our findings align with its attributes, and thus confirm its internal and external validity. However, this model, which includes ten attributes, lacks a user interface design. As previously discussed, the user interface is one of the most frequent sources of usability issues. That said, we advocate including this attribute in usability modeling. 

On top of that, usability models have been developed to effectively address user-reported issues. This is not surprising, as user acceptance testing is the final stage of the software development cycle. In light of the obtained results, however, usability testing and evaluation should begin at the design stage. In other words, our findings advocates a \textit{shift-left testing} approach. By definition, this approach emphasizes moving testing activities earlier in the development process to improve software quality through better test coverage and continuous feedback.

From a broader perspective, our results affirms the value of User-Centered Design approach on the one hand, and challenge the traditional software development lifecycle by advocating for early and continuous testing on the other. In this manner, we also acknowledge the core principle of the Agile Manifesto~\cite{agile-manifesto} regarding customer (user) collaboration.
Specifically, in line with other research, we agree that integrating users into the development process, fostering collaboration between teams, and maintaining openness and responsiveness to change are effective ways to produce user-friendly mobile applications.

\subsection{Practical implications}
Our study also informs software practitioners. 
First, we provide user interface designers with a catalog of mobile usability issues. The delivered information pertains to a variety of mobile applications, user groups, and usage contexts. In our opinion, such approach yields a comprehensive perspective on user interface design.

Second, considering the role of a software analyst, we provide a structured vocabulary for developing non-functional requirements. Additionally, one can use the selected issues to formulate measurable quality norms against the corresponding application attributes. 
If a third-party software vendor is involved, clearly defined usability requirements establish unambiguous acceptance criteria for the implemented, yet not released product.

Third, we inform software developers of hardware and network-related issues. With the need to develop resource-efficient mobile applications in mind, the source code can be debugged and checked to enable targeted fixes. Technical debt can also be reduced through targeted refactoring, which involves identifying and addressing recurring usability issues.

Fourth, we provide software testers with a ready-to-use checklist. Although our catalog is generic and structured, it can easily be adopted and extended to ensure that all relevant criteria are consistently considered during testing and evaluation. This approach, on the other hand, would enhance accountability and communication by facilitating progress tracking and clearly reporting discovered issues.

In summary, our study informs interested parties about the potential challenges and obstacles of developing mobile software. It is important to note that we strongly advocate for a data-driven approach, relying on empirical evidence rather than intuition, naive assumptions, or anecdotal experience.

\subsection{Study Limitations and Mitigation Countermeasures}
Along with the implications for theory and practice, we also recognize some limitations arising from the study settings. Although we employed a triangulation strategy involving two different research methods, each inherently carries certain limitations.

\subsubsection{Systematic Literature Review}
The first was a systematic literature review (SLR) which substantially expanded the scope of several valuable reports on mobile usability issues. 
Firstly, the search string was deliberately limited to the top-level words extracted from the research question. This approach excluded from the search string lower-level terms that were nevertheless closely related to the research theme.

Secondly, limitations also arise from the time frames and data sources. Since the search only covers studies up to a certain point in time, more recent research may not included. This results in an evidence base that does not fully reflect current developments in the field. However, this issue was addressed by selecting the fastest indexing search engine.

Thirdly, relying primarily on a single, arbitrarily selected database introduces the risk of incomplete coverage. However, to the best of our knowledge, Google Scholar offers the widest online coverage of academic content, including journal articles, reviews, and conference papers.

Fourthly, limitations also stem from the data extraction and analysis. Even when a predefined protocol is used, these stages inevitably involve subjective judgment. This is most apparent when extracting ambiguous information from different sections of a paper or reconciling discrepancies across studies. 

Fifthly, the results of documented research may be biased depending on the target user group and the specific features tested. This threat was mitigated by formulating and applying inclusion and exclusion criteria. However, their application could be misused during the reading process, which could lead to the incorrect inclusion or exclusion of certain papers. This threat was mitigated by re-checking the paper's relevance (irrelevance), after the initial decision.

\subsubsection{Interview}
An interview is a qualitative and systematic method of collecting detailed information from an individual about a specific topic.
Therefore, by nature, the gathered data may be non-generalizable and subject to bias. Nevertheless, these limitations can be mitigated to some extent. In this regard, the survey question was intentionally formulated in a general way so that we could collect responses that were not limited to specific issues tied to the unique app features. The experts understood our line of thinking, and the collected information provides a general perspective on the subject matter.

The second limitation concerns the size of the sample. 
Considering that our study involved only nine experts, issues concerning the interpretation and extrapolation of results arise. However, in our opinion, insufficient sample size is not seen to threaten the validity and generalizability. The experts did not share professional experience. They all answered on their own, without consulting each other. With an adequate education and rich professional background, we found that the data collected after the ninth interview had reached a satisfactory level of saturation, providing sufficient information to draw valid conclusions.

The third limitation stems from the use of purposive sampling. 
However, this limitation has little bearing on our study due to its qualitative nature and the scope of the research topic. Nevertheless, subjectivity and potential bias may still arise in the selection of participants, as it relies solely on the author's judgment. To mitigate this threat, the author intentionally selected more individuals than were needed for the study, anticipating that not all would be available. Additionally, each expert underwent a screening process that considered their professional experience.

\section{Conclusions}
\label{sec:conclusions}
Although usability is widely recognized as one of the most important quality factors in mobile application development, it still presents several challenges and obstacles. Identifying the sources of usability issues is undoubtedly imperative to building a user-friendly app.
To understand what constitutes a usability issue in mobile applications, it is important to consider the perspectives of users, researchers, and experts. Following this line of thinking, our study was undertaken to search for usability issues by reviewing the scholarly literature, surveying the practitioners, and investigating user online reviews. 

The results presented and discussed cover a wide range of issues that we organized into 16 categories and divided into three levels. At the first App level, usability issues penetrate the user interface design, efficiency, errors, and operability.
At the second User level, they affect cognitive load, effectiveness, ease of use, learnability, memorability, and understandability. 
At the third Resource level, usability is primarily influenced by network quality, as well as by the following hardware resources: battery, CPU, buttons, RAM, and screen size. 

In overall, this findings are in potential interest of stakeholders, including both research community and mobile software industry. While both groups are focusing on different avenues, yet they share the same concern. Specifically, researchers investigate the mobile application landscape through research and experimentation. On the other hand, practitioners are set on developing mobile applications that will satisfy users. Clearly, our results meet the needs of both groups.

Similarly, the results lay the foundation for other endeavors that can be undertaken in the future. 
Further research could involve developing a latent operationalization model for identified usability variables, which would provide a way to measure and evaluate the usability of any mobile application from the perspective of its users. Additionally, a user survey could confirm the measurement model and determine the strength and direction of relationships between specific attributes and perceived usability. 
From the perspective of software providers, our results lay the groundwork for improving the quality assurance programs, including upgrading existing reviews and audits tools and testing checklists. 



\end{document}